\documentclass[prb,showpacs,preprintnumbers,amsmath,amssymb,twocolumn]{revtex4}

\usepackage{graphicx}
\usepackage{dcolumn}
\usepackage{bm}

\preprint{submitted to Physical Review B}

\begin{document}

\title{Effect of the double-counting functional on the electronic
and magnetic properties of half-metallic magnets using the GGA+U
method}

\author{Christos Tsirogiannis}
\author{Iosif Galanakis}\email{galanakis@upatras.gr}
\affiliation{Department of Materials Science, School of Natural
Sciences, University of Paras,  GR-26504 Patra, Greece}

\date{\today}

\begin{abstract}
Methods based on the combination of the usual density functional
theory (DFT) codes with the Hubbard models are widely used to
investigate the properties of strongly correlated materials. Using
first-principle calculations we study the electronic and magnetic
properties of 20 half-metallic magnets performing self-consistent
GGA+U calculations using both the atomic-limit (AL) and
around-mean-field (AMF) functionals for the double counting term,
used to subtract the correlation part from the DFT total energy,
and compare these results to the usual
generalized-gradient-approximation (GGA) calculations. Overall the
use of AMF produces results similar to the GGA calculations. On
the other hand the effect of AL is diversified depending on the
studied material. In general the AL functional produces a stronger
tendency towards magnetism leading in some cases to unphysical
electronic and magnetic properties. Thus the choice of the
adequate double-counting functional is crucial for the results
obtained using the GGA+U method.
\end{abstract}


\maketitle

\section{Introduction}
\label{sec1}

The rapid expansion of the field of spintronics and
magneto-electronics brought magnetic materials at the nanoscale to
the center of attention of modern electronics. The spin of the
electron offers an additional degree of freedom in electronic
devices with respect to conventional electronics based on
semiconductors.\cite{ReviewSpin} The design of magnetic
nanomaterials with novel properties offers new functionalities to
future devices, and to this respect \textit{ab-initio} (also known
as first-principles) studies of the electronic structure within
density functional theory (DFT) play a crucial role allowing the
modelling of the properties of several materials prior to their
experimental growth.  Among the most studied magnetic materials
are the so-called half-metallic (HM)
magnets,\cite{ReviewHM,Review_DMS} which present metallic behavior
for the majority-spin electronic band structure and semiconducting
for the minority-spin electronic band structure. The ferromagnetic
semi-Heusler compound NiMnSb was the first material for which the
HM character was predicted and described,\cite{Groot1983} and
since then several HM compounds have been
discovered.\cite{Pickett07,FelserRev,FelserRev2} The
implementation of half-metallic magnets in devices is an active
field of research (see Ref. \onlinecite{Perspectives} for a review
of the literature).

DFT-based ab-initio electronic structure calculations using either
the local-spin-density approximation (LSDA)\cite{LSDA} or the
generalized-gradient-approximation (GGA)\cite{GGA}  for the
exchange-correlation functional are quite successful for magnetic
materials from weak to intermediate electronic correlations, but
fail for systems with strong electronic correlations. There are
two common ways to include correlations in first-principles
electronic structure calculations. The first one is the so-called
LDA+$U$ scheme, in which the local--(spin)-density approximation
(L(S)DA) of DFT is augmented by an on-site Coulomb repulsion term
and an exchange term with the Hubbard $U$ and Hund exchange $J$
parameters, respectively.\cite{JepsenPRB2010,SolovyevReview} Such
a scheme has been applied for example to Co$_2$FeSi, showing that
correlations restore the HM character of the
compound,\cite{FelserU} and to NiMnSb.\cite{ChioncelPRB06_2} When
the GGA functional is used instead of the L(S)DA the method is
usually referred to as GGA+U scheme. A more elaborate modern
computational scheme, which combines many-body model Hamiltonian
methods with DFT, is the so-called LDA+DMFT method, where DMFT
stands for Dynamical Mean-Field Theory.\cite{MinarReview,DMFT}
LDA+DMFT has been applied to several HM magnetic systems like
Co$_2$MnSi,\cite{ChioncelPRL08}
NiMnSb,\cite{ChioncelPRB03,ChioncelPRB06,ChioncelPRB10}
FeMnSb,\cite{ChioncelPRL06} Mn$_2$VAl,\cite{ChioncelMn2VAl}
VAs\cite{ChioncelVAs} and CrAs.\cite{ChioncelCrAs,ChioncelCrAs2}

In the case of both LSDA+U (GGA+U) and LDA+DMFT schemes, the
addition of the Hubbard U interaction introduces the need for a
"double-counting" correction term in the energy functional to
account for the fact that the Coulomb energy between the
correlated states is already included in the LSDA (GGA)
functional. Several double-counting schemes have been proposed in
literature,\cite{Anisimov,Anisimov2,Czyzyk} and in all proposed
schemes an averaged energy for the occupation of a selected
reference state is subtracted. Among the proposed functionals for
the double-counting term, two are most commonly used: the
so-called around-mean-field (AMF) functional and the atomic-limit
(AL) functional; the latter is also referred to in literature as
the fully localized limit (FLL) functional. The performance of
these two functionals has attracted little attention in
literature. In 2009 Ylvisaker and collaborators presented an
extensive study on the effect of the two functionals when
performing self-consistent LSDA+U calculations for several
magnetic materials.\cite{ALvsAMF} They have shown that the use of
the LSDA+U interaction term usually enhances spin magnetic
moments, but the AMF double-counting term gives magnetic states a
significantly larger energy penalty than does the AL(FLL)
functional and thus AL gives a stronger tendency to magnetism than
AMF.\cite{ALvsAMF}

\section{Motivation and computational method}
\label{sec2}

As mentioned above, ab-initio electronic structure calculations
based on the mixed LSDA+U/GGA+U schemes as well as LDA+DMFT are
widely used to study the influence of electronic correlations on
the electronic and magnetic properties of half-metallic magnets.
Thus the study of the influence of the double-counting term on the
calculated properties for these materials is extremely important
with respect to their potential use in realistic devices. The aim
of the present study is to explore the effect of both AL and AMF
functionals when performing GGA+U calculations with respect to
usual electronic band structure calculations using the GGA
functional for a wide range of half-metallic magnets (the reader
is referred to Ref. \onlinecite{ALvsAMF} for an extended
discussion on the exact formulation of the two functionals). To
achieve our goal we have employed the full-potential nonorthogonal
local-orbital minimum-basis band structure scheme
(FPLO).\cite{koepernik} For the GGA calculations we have used the
Perdew-Burke-Ernzerhof parametrization.\cite{GGA} In the case of
the GGA+U calculations the on-site Coulomb interactions for the
correlated $d$ or $p$ orbitals are introduced via the $F_0$,
$F_2$, $F_4$ and $F_6$ Slater parameters.\cite{FPLO2} For all
calculations a dense 20$\times$20$\times$20 grid in the reciprocal
space has been used to carry out the integrals and both the charge
density (up to 10$^{-6}$ in arbitrary units) and the total energy
(up to 10$^{-8}$ Hartree) have been converged in each case.

In order to cover a wide range of half-metallic magnets in a
coherent way, we have used in our calculations the ab-initio
determined Coulomb effective interaction parameters  (Hubbard $U$
and Hund exchange $J$ between localized \emph{d} or \emph{p}
electrons) calculated in Ref. \onlinecite{Gala_cRPA} using the
constrained Random Phase Approximation
(cRPA)\cite{cRPA,cRPA_2,cRPA_3,cRPA_Sasioglu} for 20 half-metallic
magnets. We should note that (i) the determination of these
parameters from experimental data is a difficult task, and (ii)
the constrained local-density approximation (cLDA), although is
the most popular theoretical
approach,\cite{cLDA1,cLDA2,cLDA3,cLDA4} it is well known to give
unreasonably large Hubbard $U$ values for the late transition
metal atoms due to difficulties in compensating for the
self-screening error of the localized electrons,\cite{cRPA_2} and
thus cRPA which does not suffer from these difficulties, although
numerically much more demanding than cLDA, offers an efficient way
to calculate the effective Coulomb interaction parameters in
solids.\cite{cRPA,cRPA_Sasioglu} We present results for all 20
half-metallic magnets studied in Ref. \onlinecite{Gala_cRPA} which
include representatives of the (i) semi-Heusler compounds like
NiMnSb, (ii) ferrimagnetic full-Heusler compounds like Mn$_2$VAl,
(iii) inverse full-Heusler compounds like Cr$_2$CoGa, (iv) usual
L$2_1$-type ferromagnetic full-Heusler compounds, (v)
transition-metal pnictides like CrAs, and finally (vi)
\emph{sp}-electron (also called \emph{d}$^0$) ferromagnets like
CaN. We have used the lattice parameters presented in Table 1 of
Ref.  \onlinecite{Gala_cRPA}. The Slater parameters entering the
FPLO method are connected to the Hubbard parameter $U_{LDA+U}$ and
to the Hund exchange $J$ presented in Table II of Ref.
\onlinecite{Gala_cRPA} for the correlated $p$-states though the
relations
\begin{equation}
F_0=U_{LDA+U},\:F_2=5\times J,\:F_4=F_6=0, \label{p}
\end{equation}
and for the correlated $d$-states
\begin{equation}
F_0=U_{LDA+U},\:\frac{F_2+F_4}{14}=J,\:\frac{F_4}{F_2}=0.625,\:F_6=0.
\label{d}
\end{equation}
We should note here that $U_{LDA+U}$ is an effective parameter
depending both on the on-site intra-orbital Coulomb repulsion
between electrons occupying the same orbital and on-site
inter-orbital Coulomb repulsion between electrons occupying
orbitals of the same $\ell$ character but different $m_\ell$
value. Thus, our study covers a wide range of half-metallic
magnets allowing for a deeper understanding of the behavior of the
AL and AMF double-counting functionals in the GGA+U calculated
electronic and magnetic properties of different HM magnetic
systems.

\section{Results and Discussion}\label{sec3}

\subsection{Binary Compounds}
\label{sec3a}

We will start the presentation of our results from the binary
compounds. There are two families of half-metallic binary
compounds. The first includes the so-called \emph{sp}-electron
ferromagnets (also known as
\emph{d}$^0$-ferromagnets).\cite{Geshi,Laref} These compounds
adopt the rocksalt cubic structure and have no transition-metal
atoms in their chemical formula. We consider the nitrides and the
carbides (CaN, SrN, SrC, and BaC) since they have the largest
calculated Curie temperatures among the studied \emph{sp}-electron
ferromagnets.\cite{N1,N2,N3,N4,C1,C2,C3} Their total spin magnetic
moment in units of $\mu_B$ equals $8-Z_{\mathrm{t}}$, where
Z$_{\mathrm{t}}$ is the total number of valence electrons in the
unit cell; this behavior is known as Slater-Pauling
(SP).\cite{Slater,Pauling}  A detailed discussion on the origin of
this rule and its connection to the half-metallicity can be found
in Ref. \onlinecite{Laref}. The usual GGA calculations produced
for all four studied compounds a half-metallic state with total
spin magnetic moments of 1 $\mu_B$ for the nitrides and 2 $\mu_B$
for the carbides. The results are gathered in Table \ref{table1}.
The spin moment is carried mainly by the N and C atoms. Our
calculated GGA results are similar to the GGA ones derived with
the full-potential linearized augmented plane-wave (FLAPW) method
as implemented in the \texttt{FLEUR}\cite{Fleur} code in Ref.
\onlinecite{Gala_cRPA}. The use of the AMF within the GGA+U scheme
leaves intact both the calculated spin magnetic moments and
density of states (DOS) with respect to GGA calculations (we do
not present the DOS since they are similar to the ones presented
in literature). On the contrary the use of the AL functional has a
tremendous effect on the calculated results. It produces an
unreasonable and unphysical charge transfer from the Ca(Sr,Ba)
atoms to the N(C) atoms resulting to huge values of the
atom-resolved spin moments. This state is obviously an artifact of
the method. We cannot explain the origin of this behavior but
starting form various configurations all calculations involving
the AL functional converged to the same results and thus the
breakdown of the AL should be attributed to its characteristics.

\begin{table}
\caption{Atom-resolved and total spin magnetic moment per formula
unit for the XY binary compounds. Results have been obtained
within the FPLO method\cite{koepernik} using the GGA functional
for the exchange interaction potential\cite{GGA} and the GGA+U
scheme employing both the atomic-limit (AL - also known as
fully-localized-limit FLL) and the around-mean-field (AMF)
functionals for the double counting term. Values for the on-site
Coulomb and exchange parameters are the ab-initio determined ones
within the constrained Random-Phase-Approximation (cRPA) in Ref.
\onlinecite{Gala_cRPA}. Lattice constants are the ones presented
in Table 1 in the later reference. Note that for the compounds
which do not contain transition metal atoms (known as
d$^0$-ferromagnets) GGA+U within the AL functional gives
unrealistic results.}
\begin{ruledtabular}
\begin{tabular}{llrrrrrr}
 Comp. & Functional & $m_{\textrm{X}}$ &
$m_{\textrm{Y}}$  & $m_{\textrm{total}}$ \\
\hline CaN        &GGA& -0.065 & 1.065 &1.000  \\
         & GGA+U (AL) & 11.836 & -4.836 & 7.000\\
       &   GGA+U (AMF) & -0.065 & 1.065 &1.000 \\
\hline SrN        &GGA& -0.072 & 1.072 &0.999  \\
         & GGA+U (AL) & 16.363 & -9.362 & 7.000\\
       &   GGA+U (AMF) & -0.072 & 1.072 & 0.999\\
\hline SrC        &GGA& -0.004 & 2.004 & 1.999 \\
         & GGA+U (AL) & 15.578 & -9.578 & 6.001\\
       &   GGA+U (AMF) &  -0.004&  2.004& 1.999\\
\hline BaC        &GGA&  0.057& 1.943 &2.000  \\
         & GGA+U (AL) &  3.261& 0.378&3.999 \\
       &   GGA+U (AMF) &  0.057&  1.943& 2.000\\
\hline VAs        &GGA&  2.427& -0.427 & 2.000 \\
         & GGA+U (AL) &  2.415&  -0.415& 2.000\\
       &   GGA+U (AMF) &  2.151&  -0.151& 2.000\\
\hline CrAs       &GGA&  3.614& -0.614 &3.000  \\
         & GGA+U (AL) &  3.880&  -0.880& 3.000\\
       &   GGA+U (AMF) &  3.541&  -0.541& 3.000\\
\hline MnAs       &GGA&  4.173& -0.311 & 3.862 \\
         & GGA+U (AL) &  4.476&  -0.476& 3.999\\
       &   GGA+U (AMF) &  3.979&  -0.326& 3.652\\
\end{tabular}
\end{ruledtabular}
\label{table1}
\end{table}

The second family of binary compounds under study are the binary
VAs, CrAs, and MnAs transition metal pnictides. The first
observation of such a compounds being half-metal was made in 2000
when Akinaga and his collaborators managed to grow multilayers of
CrAs/GaAs.\cite{Akinaga} CrAs was found to adopt the zincblende
structure of GaAs and was predicted to be a half-metal with a
total spin magnetic moment of 3 $\mu_B$ in agreement with
experiments.\cite{Akinaga} Several studies followed this initial
discovery, and electronic structure calculations have confirmed
that also similar binary XY compounds, where X is an early
transition-metal atom and Y an \emph{sp} element, should be
half-metals and the total spin magnetic moment follows a SP rule
similar to d$^0$-ferromagnets being now equal to
$Z_{\mathrm{t}}-8$.\cite{Mavropoulo03,ReviewZB}

In Table \ref{table1} we gathered all the calculated spin magnetic
moments. GGA gives a half-metallic state for VAs and CrAs, while
for MnAs the Fermi level is slightly above the minority-spin
energy gap and the total spin magnetic moment slightly smaller
than the ideal value of 4 $\mu_B$ for half-metallicity to occur.
These results have been largely discussed in
literature.\cite{Mavropoulo03,ReviewZB} For both VAs and CrAs,
GGA+U self-consistent calculations yield a half-metallic state
within both AL and AMF functionals with the same total spin
magnetic moment but with substantial variations of the
atom-resolved spin magnetic moments. For MnAs the use of AL
functional leads to a half-metallic state contrary to AMF for
which the Fermi level is above the minority-spin gap. Overall AL
leads to larger absolute values of the atomic spin moments with
respect to GGA while AMF leads to smaller values. This behavior of
the atomic spin magnetic moments confirms the conclusion in Ref.
\onlinecite{ALvsAMF} that AMF gives the magnetic state a large
energy penalty with respect to AL.

\begin{figure}[t]
\begin{center}
\includegraphics[width=\columnwidth]{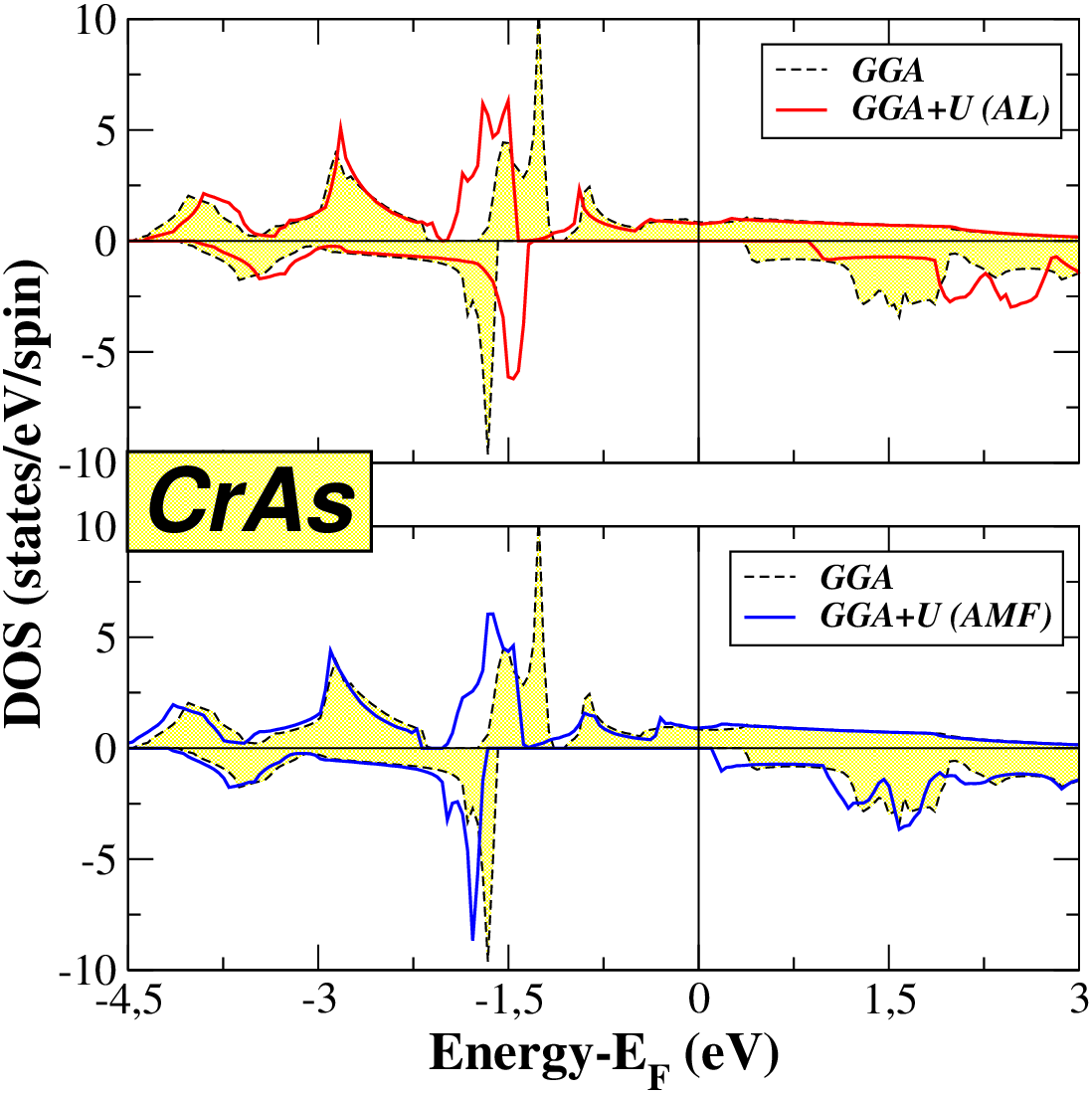}
\end{center}
\vspace*{-0.6 cm}\caption{(color online) Total density of states
(DOS) as a function of the energy for the CrAs compound within the
GGA+U method using both the atomic-limit (AL) and the
around-mean-field (AMF) functionals for the double-counting term.
GGA+U results are compared to the GGA calculated DOS. The zero in
the energy axis has been set to the Fermi level.
Positive(negative) values of the DOS correspond to the
majority(minority)-spin electrons.} \label{fig1}
\end{figure}

Since DOS present similar trends between the three transition
metal binary compounds, we present in Fig. \ref{fig1} the
calculated DOS per formula unit for CrAs. In the upper panel we
compare the GGA+U calculated DOS within the AL functional to the
usual GGA calculated DOS, and in the lower panel we present a
similar graph for the AMF case. In the presented energy Cr DOS
dominates. GGA produces a large minority-spin gap with a large
exchange splitting between the occupied majority-spin bands and
the unoccupied minority-spin bands and thus strong tendency to
magnetism manifested also by the large ($\sim$3.6 $\mu_B$) Cr spin
moment. The use of the AL double-counting functional in the GGA+U
calculations lead to an almost rigid shift of the minority spin
DOS towards higher energies, while in the majority-spin DOS only
the double-degenerate $e_g$ states at about -1.5 eV move lower in
energy (see Ref. \onlinecite{Mavropoulo03} for a discussion of the
character of the bands). In the case of AMF the majority-spin band
structure shows a similar behavior with respect to the GGA results
as the AL case. But in the minority-spin band structure the
tendency is the opposite now. Since AMF does not favor magnetism
as strongly as AL, the minority-spin band structure now presents
an almost rigid shift towards lower energy values. These finding
also explain the behavior of the MnAs compound. In the case of AL
the minority-spin band structure moves towards higher energy
values and the Fermi level now moves within the gap and
half-metallicity appears.

\subsection{Heusler compounds} \label{sec3a}

Heusler compounds are a huge family of intermetallic compounds
presenting various types of electronic and magnetic
behaviors.\cite{landolt,landolt2} Several among them are
half-metallic ferromagnets/ferrimagnets/antiferromagnets and are
of particular interest due to their very high Curie temperatures,
which usually exceed 1000 K, making them ideal for
applications.\cite{Perspectives} There are four main families of
Heusler compounds: (i) the semi-Heuslers also known as
half-Heuslers like NiMnSb which have the chemical type XYZ with X
and Y being transition metal atoms, (ii) the usual full Heuslers
like Co$_2$MnSi with the chemical type X$_2$YZ where the valence
of X is larger than the valence of Y and the two X atoms are
equivalent, (iii) the quaternary Heuslers like (CoFe)MnSi which
present similar properties with the full-Heuslers, and finally
(iv) the so-called inverse-Heuslers, like Cr$_2$CoGa which have
also the chemical type X$_2$YZ but now the valence of X is smaller
than the valence of Y and due to the change of the sequence of
atoms in the unit cell the two X atoms are no more
equivalent.\cite{landolt,landolt2} We present results for all
families of compounds with the exception of quaternary-Heuslers
which present similar behavior to the full-Heuslers and for which
no Hubbard parameters have been derived in Ref.
\onlinecite{Gala_cRPA}.

\begin{figure}[t]
\begin{center}
\includegraphics[width=\columnwidth]{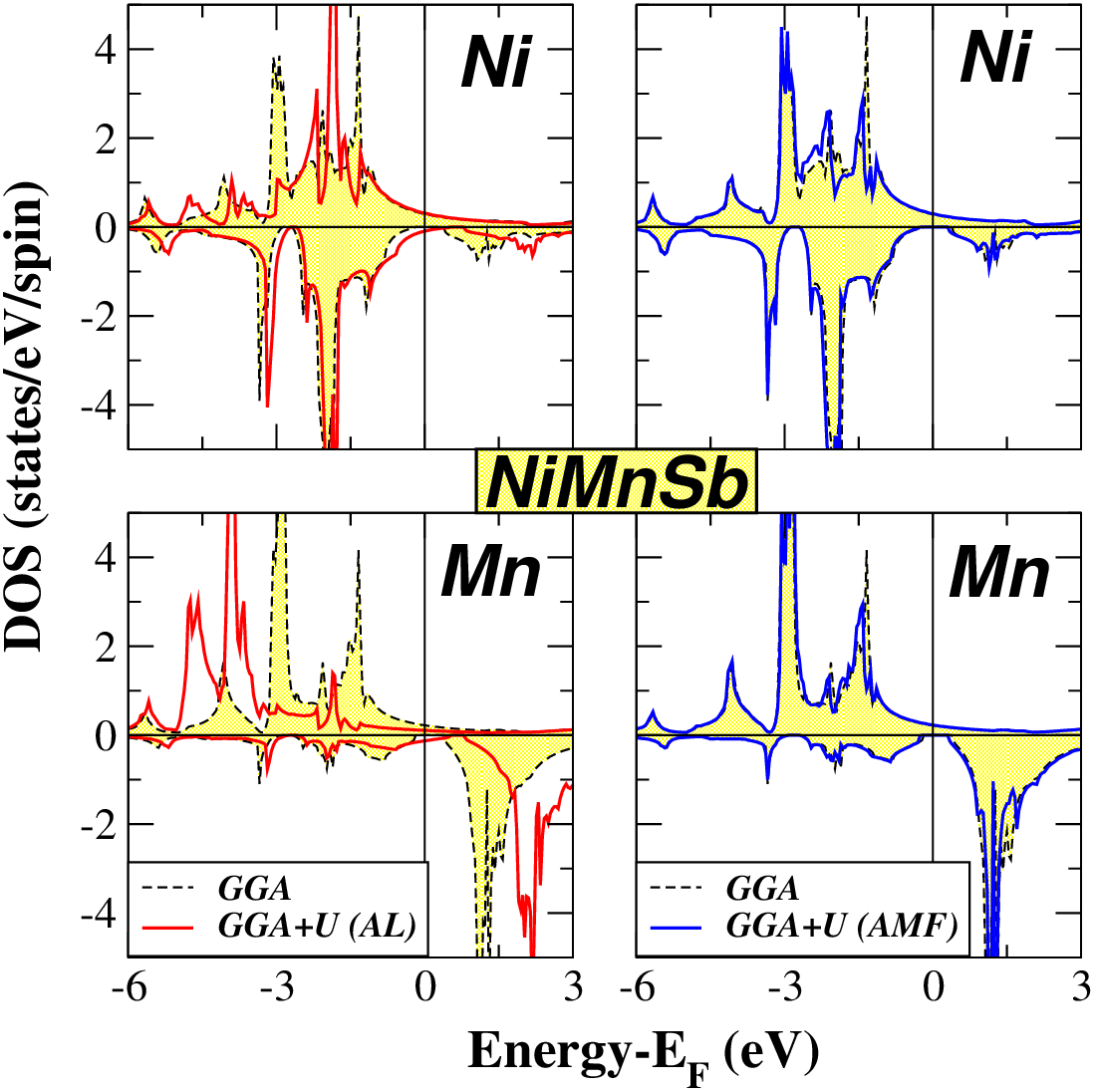}
\end{center}
\vspace*{-0.6 cm}\caption{(color online) Ni and Mn atom-resolved
DOS in NiMnSb. Details as in Fig. \ref{fig1}.} \label{fig2}
\end{figure}

\subsubsection{Semi-Heuslers}

The first family of Heusler compounds for which we will present
results are the semi-Heuslers. The first compound that was
predicted to be a half-metal was actually a semi-Heusler,
NiMnSb.\cite{Groot1983} Their total spin magnetic moment follows
also a SP rule being $Z_t$-18 (for an extended discussion see Ref.
\onlinecite{Gala02a}). In Table \ref{table2} we have gathered our
calculated spin magnetic for all studied cases and for three
compounds FeMnSb, CoMnSb and NiMnSb (note that for FeMnSb we were
not able to converge the GGA+U calculations using the AMF
functional). As we move from one compound to the other, the total
number of valence electrons increases by one and so does the GGA
calculated total spin magnetic moment. Mn atoms in all case posses
a large value of spin magnetic moment which starts from $\sim$3.4
$\mu_B$ in FeMnSb and exceeds 4 $\mu_B$ in NiMnSb. As we increase
the total number of valence electrons the spin magnetic of the X
atoms also increases being $\sim$-1.3 $\mu_B$ for Fe, -0.34
$\mu_B$ for Co and 0.14 $\mu_B$ for Ni in the corresponding
compounds. The GGA calculated DOS, presented in Fig. \ref{fig2}
for NiMnSb, has been studied in detail in literature and it is
mainly characterized by the large exchange splitting between the
occupied majority-spin and the unoccupied minority-spin $d$-states
at the Mn site which together with the very small weight of the
occupied minority-spin states are responsible for the large Mn
spin magnetic moments. This feature is common for all three
studied compounds and has been already observed in
literature.\cite{Gala02a,FeMnSb_dos,CoMnSb_dos}

\begin{table}
\caption{Similar to Table \ref{table1} for the semi-Heusler
compounds crystallizing in the C1$_b$ lattice structure having the
XYZ chemical formula.}
\begin{ruledtabular}
\begin{tabular}{llrrrrrr}
 Comp. & Functional & $m_{\textrm{X}}$ &
$m_{\textrm{Y}}$ &  $m_{\textrm{Z}}$ &  $m_{\textrm{total}}$ \\
\hline FeMnSb      &GGA&-1.279 &3.374 & -0.095 & 2.000 \\
        &   GGA+U(AL) &-2.274& 4.559 &  0.042 & 2.327\\
\hline CoMnSb     &GGA& -0.340&3.568 &-0.227 &  3.000 \\
         & GGA+U (AL) &-1.264 & 4.520&-0.186 &  3.068\\
       &   GGA+U (AMF) &-0.358& 3.535 &-0.177 &  3.000\\
\hline NiMnSb    & GGA& 0.143 &4.031 &-0.174 &  3.999 \\
         & GGA+U (AL) & -0.087 & 4.553&-0.301 &  4.164\\
       &   GGA+U (AMF) &  0.144& 4.026 &-0.171 &  3.999\\
\end{tabular}
\end{ruledtabular}
\label{table2}
\end{table}

The self-consistent GGA+U calculations using the AMF functional
for the CoMnSb and NiMnSb compounds produced a similar picture to
the GGA calculations. The total spin magnetic moment, as shown in
Table \ref{table2} remains identical to the GGA case and the
atom-resolved spin magnetic moments only scarcely changed. This is
also reflected on the Ni and Mn resolved DOS for NiMnSb in Fig.
\ref{fig2} where the GGA+U within AMF calculated DOS is almost
identical to the GGA calculated DOS. The effect of the use of the
AL functional is more drastic. As shown in Fig. \ref{fig2} GGA+U
within the AL functional compared to usual GGA leads to large
modifications of the DOS of the transition metal atoms. In the
case of Mn, the exchange splitting between the occupied
majority-spin and the unoccupied minority-spin states increase
considerably. In the majority-spin band structure of Mn, the
occupied states move lower in energy and as a result they are no
more in the same energy with the Ni majority-spin states. This
leads to a weaker hybridization between the $d$ states of Ni and
Mn atoms and in the Ni DOS the width of the majority bands becomes
smaller as a result of the weaker hybridization effects. Almost
all the weight of the  occupied minority-spin band structure is
located at the Ni atoms while almost all unoccupied minority-spin
states are located at the Mn atom. Thus there is almost no
hybridization between the minority-spin $d$-states of Ni and Mn
and the former are not affected by the shift of the later to
larger energy values being identical to the GGA case.   These
changes in DOS are also reflected on the spin magnetic moments in
Table \ref{table2}. The larger tendency to magnetism within AL
compared to AMF leads to slightly larger total spin magnetic
moments which deviate from the ideal integer values of the SP rule
and the Fermi level is located slightly below the minority-spin
energy gap. In the case of FeMnSb and CoMnSb, both the absolute
values of the Fe(Co) and Mn spin magnetic moments increase by
about 1 $\mu_B$ almost cancelling each other. In the case of
NiMnSb the variations  in the atomic spin magnetic moments are
considerably smaller since almost all Ni $d$-states are occupied
in all studied cases. But even for NiMnSb the Mn spin moment
increases by $\sim$0.5 $\mu_B$ and the Ni spin moment decrease by
about 0.3 $\mu_B$. The Sb atoms in all cases also present changes
in their atomic spin magnetic moments between the various
calculations although these variations are considerably smaller
than for the transition metal atoms.

\begin{table}
\caption{Similar to Table \ref{table1} for the full-Heusler
compounds crystallizing in the L2$_1$ lattice structure having the
X$_2$YZ chemical formula.}
\begin{ruledtabular}
\begin{tabular}{llrrrrrr}
 Comp. & Functional & $m_{\textrm{X}}$ &
$m_{\textrm{Y}}$ &  $m_{\textrm{Z}}$ &  $m_{\textrm{total}}$ \\
\hline Mn$_2$VAl  &GGA& -1.670 &1.227 & 0.113 & -1.999 \\
         & GGA+U (AL) & -4.102 & 3.169& 0.483 & -4.551\\
       &   GGA+U (AMF) & -1.798& 1.527 & 0.079 & -1.990\\ \hline
 Mn$_2$VSi  &GGA& -0.801 &0.557 & 0.060 & -0.985 \\
   & GGA+U (AL) & -2.248 & 2.759& 0.376 & --1.362\\
\hline Co$_2$CrAl&GGA& 0.737 &1.684 & -0.160 & 2.999 \\
        &   GGA+U (AMF) & 0.965& 1.222 & -0.153 & 3.000\\
\hline Co$_2$CrSi  &GGA& 0.934 &2.242 &-0.111 &  4.000 \\
         & GGA+U (AL) & 0.871 & 2.525&-0.267 &  4.000\\
       &   GGA+U (AMF) & 0.890& 2.187 &0.031  &  4.000\\
\hline Co$_2$MnAl & GGA& 0.673 &2.910 &-0.231 &  4.025 \\
         & GGA+U (AL) &  1.381 & 4.176&-0.501 &  6.438\\
       &   GGA+U (AMF) &  1.048& 1.998 &-0.096 &  3.999\\
\hline Co$_2$MnSi & GGA& 0.972 &3.195 &-0.140 &  4.999 \\
         & GGA+U (AL) &  0.732 & 3.954&-0.338 &  5.080\\
       &   GGA+U (AMF) &  0.987& 3.020 &-0.004 &  5.000\\
\hline Co$_2$FeAl & GGA& 1.163 &2.870 &-0.203 &  4.999 \\
         & GGA+U (AL) &  1.188 & 3.326&-0.520 &  5.177\\
       &   GGA+U (AMF) &  1.216& 2.673 &-0.105 &  4.999\\
\hline Co$_2$FeSi & GGA& 1.327 &2.926 &-0.042 &  5.539 \\
         & GGA+U (AL) &  1.375 & 3.450&-0.201 &  5.999\\
\end{tabular}
\end{ruledtabular}
\label{table3}
\end{table}

\subsubsection{Full-Heuslers}

The second family of Heusler compounds which may present
half-metallicity are the so-called usual full-Heuslers
crystallizing in the cubic L2$_1$ structures. Half-metallicity can
be combined either with the appearance of ferrimagnetism, when the
X atoms in the X$_2$YZ is the Mn one, or with ferromagnetism when
X is Co. In all cases the total spin magnetic moment in $\mu_B$
follows a SP rule being $Z_t$-24.\cite{Gala02b} In Table
\ref{table3} we have gathered the calculated spin magnetic moments
for all studied compounds with both GGA and GGA+U methods using
both the AL and AMF double-counting functional in the later case.
When one case is missing in the table, this is due to the fact
that we were not able to get convergence irrespectively of the
starting input which we have used.

First, we will discuss our results on the half-metallic
ferrimagnetic Mn$_2$VAl and Mn$_2$VSI compounds where the total
spin magnetic moments is negative since the total number of
valence electrons is less than 24. Moreover the Mn spin magnetic
moments are antiferromagnetically coupled to the V spin moments
due to their small distance.\cite{Gala02b} In the case of
Mn$_2$VAl, GGA+U calculations within AMF produced similar spin
moments and DOS to the GGA case; we were not able to converge
GGA+U within AMF for the Mn$_2$VSi compound. For both compounds
the use of AL double-counting functional produced unphysical
results similar to the case of $d^0$-ferromagnets in Sec.
\ref{sec3a}. The use of AL tripled, with respect to the GGA case,
the absolute values of the spin magnetic moments of the transition
metal atoms in Mn$_2$VA; in the case of V in Mn$_2$VSi the
increase is almost 600\% . Thus the use of AL for the
half-metallic ferrimagnetic Heusler compounds obviously is
inadequate.

\begin{figure}[t]
\begin{center}
\includegraphics[width=\columnwidth]{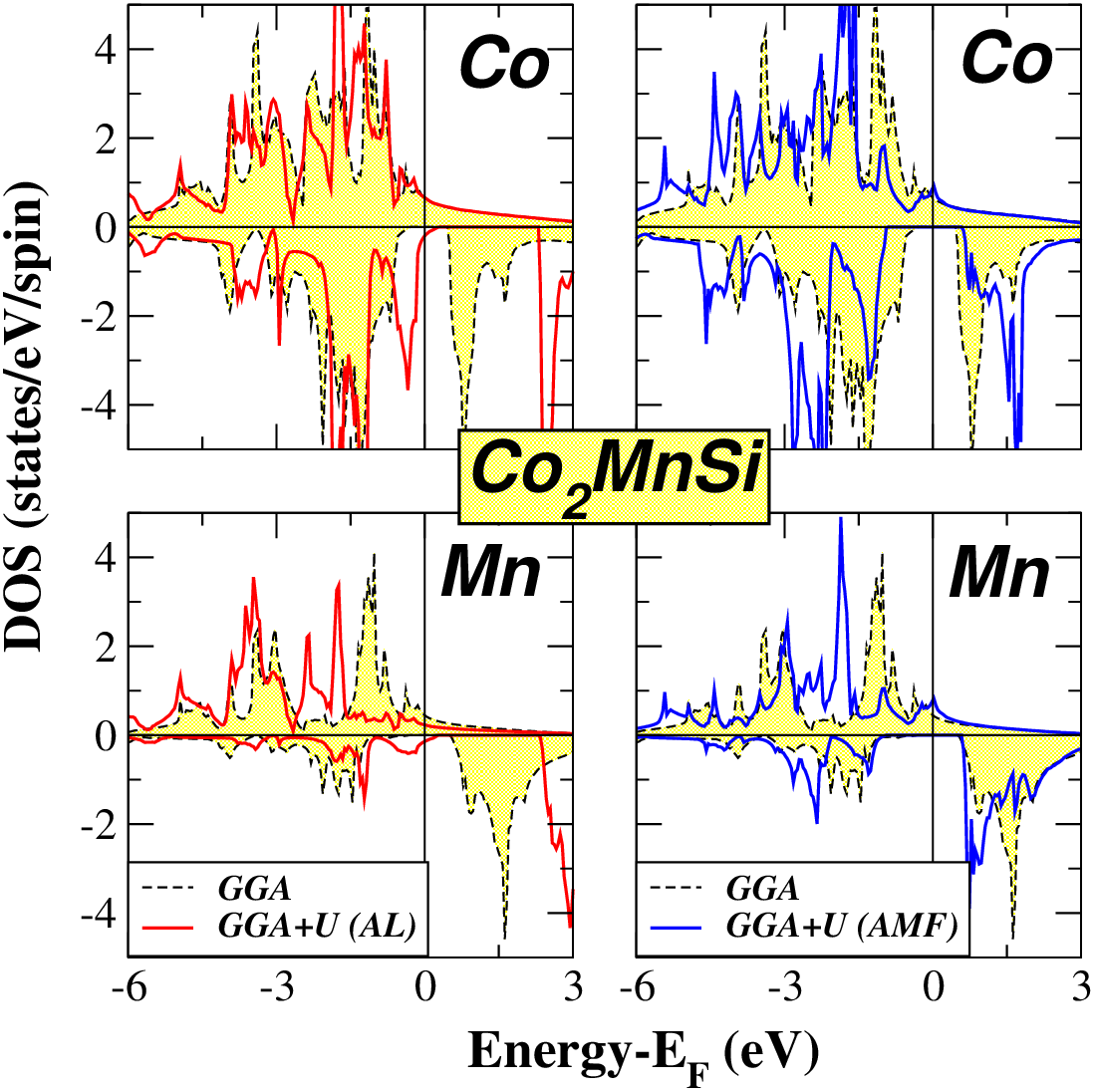}
\end{center}
\vspace*{-0.6 cm}\caption{(color online) Co and Mn atom-resolved
DOS in Co$_2$MnSi. Details as in Fig. \ref{fig1}.} \label{fig3}
\end{figure}

In the case of ferromagnetic full-Heuslers containing Co the
effect of using both AMF and AL on the calculated electronic and
magnetic properties is more complex than in all the previously
studied cases. When Y is Cr (Co$_2$CrAl and Co$_2$CrSi) both AL
and AMF yielded a perfect half-metallic state with the total spin
magnetic moment being equal to the ideal values predicted by the
SP rule as shown in Table \ref{table3}. When Y is Mn (Co$_2$MnAl
and Co$_2$MnSi) AMF produced a half-metallic states and both
atom-resolved and total spin magnetic moments where close to the
GGA case, but AL led to a considerable increase of the Mn spin
moment similarly to the semi-Heuslers. The increase of the Mn spin
moment within AL led to an increase also of the total spin
magnetic moment which is only 0.80 $\mu_B$ for Co$_2$MnSi but
reaches the $\sim$1.4 $\mu_B$ for Co$_2$MnAl. When Y is Fe
(Co$_2$FeAl and Co$_2$FeSi) the behavior of the spin moments with
respect to the GGA results is similar within both AL and AMF to
the case where Y is Mn. Moreover in the case of Co$_2$FeSi which
is not half-metallic within GGA, the use of GGA+U combined with AL
leads to a total spin magnetic moment of 6 $\mu_B$ and to a
half-metallic state as shown also in Ref. \onlinecite{FelserU}.
Although the GW scheme\cite{BlugelGW} produced similar results to
the GGA+ calculations, correlations in this materials are still an
open issue since recent results by Meinert and collaborators show
that a self-consistent calculation fixing the total spin magnetic
moment to 6 $\mu_B$ reproduces more accurately the position of the
band with respect to available experimental
data.\cite{MeinertCo2FeSi}

\begin{figure}[t]
\begin{center}
\includegraphics[width=\columnwidth]{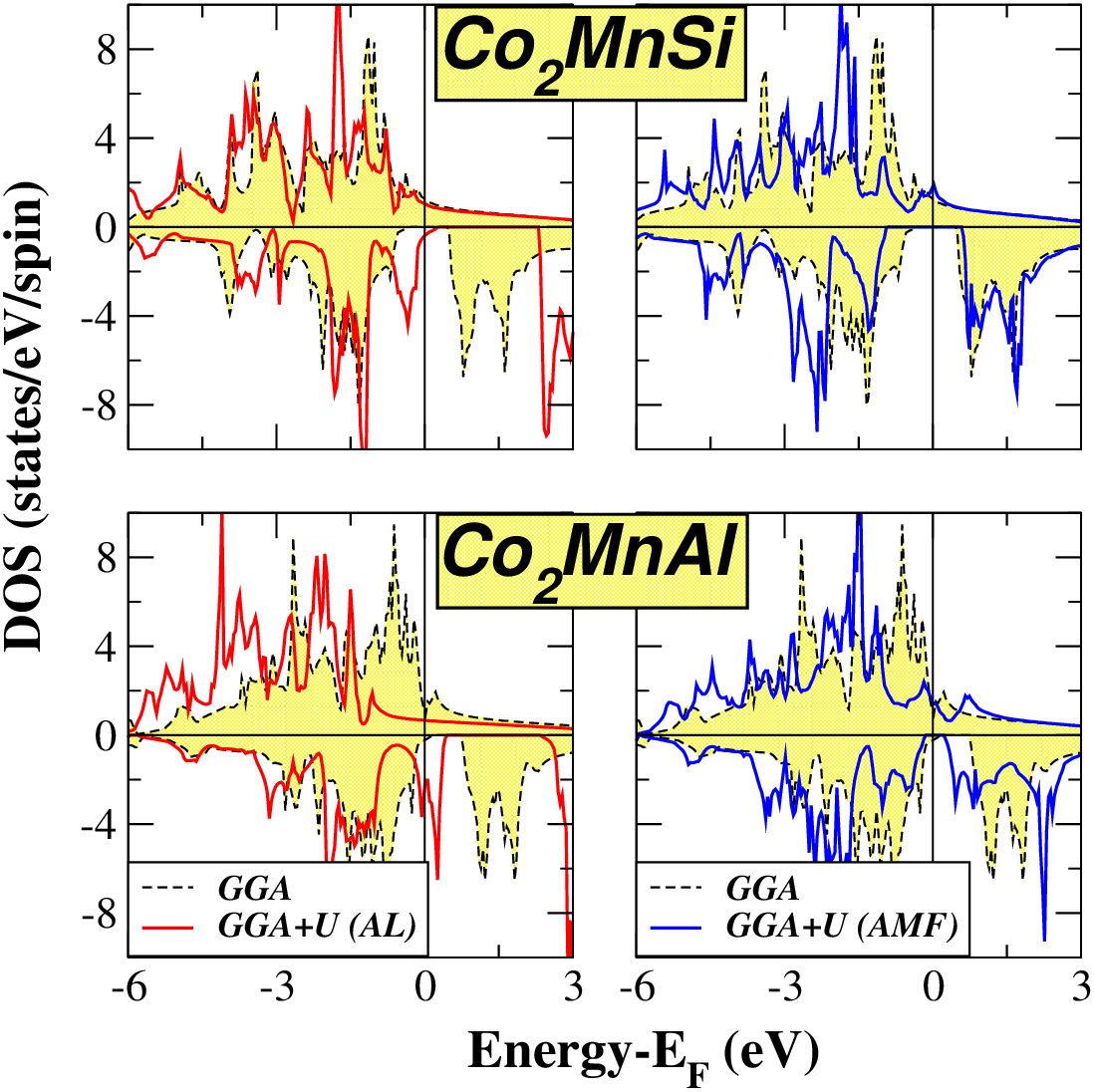}
\end{center}
\vspace*{-0.6 cm}\caption{(color online) Total DOS per formula
unit for the Co$_2$MnSi and Co$_2$MnAl compounds. Details as in
Fig. \ref{fig1}.} \label{fig4}
\end{figure}

To understand the behavior of the spin moments we have to examine
in detail the behavior of the DOS. Since the trends when either
AMF or AL is employed are similar for all six ferromagnetic
Co-based full-Heuslers under study, we will use Co$_2$MnSi as an
example and in Fig. \ref{fig3} we present the Co and Mn resolved
DOS. When the GGA+U combined with AMF is used (left panel) there
is a significant change in the DOS unlikely all other families of
half-metallic compounds discussed above. AMF enhances the tendency
to magnetism with respect to GGA. For the Mn atom the occupied
majority spin states shift lower in energy but the minority-spin
energy gap in the Mn DOS remains unchanged. Through hybridization
also the Co majority-spin DOS shifts lower in energy and so do
also the occupied Co minority-spin states. This leads to an
increase of the energy gap in the Co minority-spin band structure.
Since as explained in Ref. \onlinecite{Gala02b} Co atoms present
in usual GGA calculations a much smaller gap than the Mn atoms and
thus determine the energy gap in the total DOS, the opening of the
former further stabilizes the half-metallic state.

The GGA+U method combined with AL even further enhances the
tendency to magnetism with respect to AMF as concluded in Ref.
\onlinecite{ALvsAMF}. In the case of Mn atoms the exchange
splitting between the occupied majority-spin and unoccupied
minority-spin states is greatly enhanced as for Mn in NiMnSb and
thus the energy gap becomes much larger. As a side-effect some
weight in the minority-spin band structure appears just below the
Fermi level. Thus although with respect to the GGA case, AL opens
the gap the Fermi level is located close to the left edge of the
gap instead of the middle in the GGA case. Co DOS follows through
hybridization the behavior of the Mn $d$-states and the gap is now
also much larger but the occupied minority-spin states move closer
to the Fermi level which now just crosses the states just below
the low-energy edge of the gap and the total spin magnetic moment
within AL is slightly larger than the ideal value of 5 $\mu_B$.

In the case of Co$_2$MnAl the change in the spin magnetic moments
is larger within both AL and AMF functionals. As shown in Fig.
\ref{fig4}, although we just change Al for Si in Co$_2$MnSi, the
AMF DOS shows a different tendency with respect to the energy gap.
The exchange splitting between occupied majority and unoccupied
minority spin states is smaller, and within AMF the gap is smaller
than within GGA showing the contrary tendency to Co$_2$MnSi where
AMF produced a larger gap with respect to GGA. For Co$_2$MnAl
within usual GGA the Fermi level is close to the left edge of the
gap while for Co$_2$MnSi it is located at the middle of the gap.
Thus in the case of AL based calculations the shift of the Co
occupied minority spin states towards higher energies for
Co$_2$MnAl, discussed just above also for Co$_2$MnSi, leads to the
loss of the half-metallicity since now the Fermi level crosses the
occupied minority-spin states. The other Al-based Heuslers
(Co$_2$CrAl and Co$_2$FeAl) exhibit within GGA a DOS around the
minority-spin energy gap similar to Co$_2$MnSi and not Co$_2$MnAl
and thus the increase in their total spin magnetic moment within
AL is much smaller than for Co$_2$MnAl.

\subsubsection{Inverse-Heuslers}

\begin{table}
\caption{Similar to Table \ref{table1} for the full-Heusler
compounds crystallizing in the inverse $XA$ lattice structure
having the Cr$_2$YZ chemical formula, where the two Cr atoms
occupy sites of different symmetry (see text).}
\begin{ruledtabular}
\begin{tabular}{llrrrrrrr}
 Comp. & Functional & $m_{{\textrm{Cr}}^A}$ & $m_{{\textrm{Cr}}^B}$ &
$m_{\textrm{Y}}$ &  $m_{\textrm{Z}}$ &  $m_{\textrm{total}}$ \\
\hline Cr$_2$FeGe &GGA&-1.454 &1.753& -0.313&  0.027 & 0.012 \\
        &   GGA+U(AL) &-4.162&4.672& -1.054&  0.551 & 0.006\\
\hline Cr$_2$CoGa  &GGA& -2.680&1.973&0.379  &-0.014 &  0.069 \\
         & GGA+U (AL) &-4.860 & 4.137&1.160&-0.082 &  0.520
\end{tabular}
\end{ruledtabular}
\label{table4}
\end{table}

The last family of potential half-metallic Heusler compounds ar
the so-called inverse Heusler compounds.\cite{GSP} Among these
half-metals the most interesting are the so-called
fully-compensated ferrimagnets (also known as half-metallic
antiferromagnets) like Cr$_2$FeGe and Cr$_2$CoGa. These materials
are of special interest since they combine half-metallicity to a
zero total net magnetization and thus are ideal for
spintronic/magnetoelectronic devices due to the vanishing external
stray fields created by them.\cite{Leuken} We should note that
films of Cr$_2$CoGa have been grown experimentally\cite{Cr2CoGa}
and this compounds has been predicted to exhibit extremely large
Curie temperature.\cite{APL11} As shown in Table \ref{table4} GGA
yields for both Cr$_2$FeGe and Cr$_2$CoGa compounds a total spin
magnetic moment close to  zero (for an extended discussion on the
half-metallic inverse Heuslers see Ref. \onlinecite{GSP}). Note
that we have two inequivalent Cr atoms in these compounds denoted
by the superscripts  $A$ and $B$ in Table \ref{table4}. We were
not able to converge the GGA+U self-consistent calculations using
the AMF double-counting functional. For the AL functional although
the total spin magnetic moment stays close to zero, the absolute
values of the Cr spin magnetic moments are about doubled leading
to an unphysical situations. Thus for these materials the use of
GGA+U combined with AL is not able to produce a reasonable
description of the electronic structure as was also the case for
the semi-Heuslers and the ferrimagnetic full-Heuslers.

\section{Conclusions}
\label{sec4}

We have studied the electronic and magnetic properties of 20
half-metallic magnets performing self-consistent GGA+U
calculations using both the atomic-limit (AL) and
around-mean-field (AMF) functionals for the double counting term
and compared them to the usual GGA calculations. Overall the use
of AMF produced results similar to the usual GGA calculations. The
effect of AL was diversified depending on the studied material. In
the case of $d^0$-ferromagnets, semi-Heuslers, ferrimagnetic
full-Heuslers and inverse Heuslers the use of AL leads to
unrealistic electronic and magnetic properties of the studied
compounds and thus its use is not justified. On the other hand in
the case of transition-metal binary compounds and usual
ferromagnetic full-Heusler compounds the use of AL enhanced the
tendency towards magnetism with respect to both GGA and GGA+U
combined with AMF. Depending on the position of the Fermi level,
there were cases like MnAs and Co$_2$FeSi for which AL produced a
half-metallic state contrary to GGA and GGA+U combined with AMF,
cases like VAs, CrAs and Co$_2$CrSi where all three methods
produced a half-metallic state, and cases like Co$_2$MnAl,
Co$_2$MnSi and Co$_2$FeAl where the use of AL led to the loss of
half-metallicity.

Methods based on the combination of the usual density functional
theory (DFT)-based codes and of the Hubbard $U$- Hund's exchange
$J$ are widely used to investigate the properties of strongly
correlated materials. Our results suggest that  especially in the
case of half-metallic magnets the choice for the double counting
functional used to subtract the part from the DFT total energy,
which is associated to the Coulomb repulsion between the
correlated orbitals, plays a decisive role on the obtained
results. Thus the study of correlations in half-metallic magnets
is still an open issue and further studies are needed to establish
the predictive power of methods based on the $U$ and $J$
parameters like GGA+U or LDA+DMFT methods.

\end{document}